\documentstyle[12pt,aasms4]{article}

\begin{document}

\title{Compact Nuclei in Moderately Redshifted Galaxies}
\author{Vicki L. Sarajedini}
\affil{Steward Observatory, University of Arizona, Tucson, AZ 85721}
\authoremail{vicki@as.arizona.edu}
\author{Richard F. Green}
\affil{NOAO\footnote{The National Optical Astronomy Observatories
are operated by the Association of Universities for Research in
Astronomy, Inc., under Cooperative Agreement with the National
Science Foundation.}, P.O. Box 26732, Tucson, AZ 85726}
\authoremail{green@noao.edu}
\author{Richard E. Griffiths and Kavan Ratnatunga}
\affil{Department of Physics and Astronomy, Johns Hopkins 
University, Baltimore, MD 21218}
\authoremail{griffith@mds.pha.jhu.edu, kavan@mds.pha.jhu.edu}

\begin{abstract}

The Hubble Space Telescope WFPC2 is
being used to obtain high-resolution images ($\le$
0.15'' FWHM) in the V (F606W) and I (F814W) bands for several thousand
distant galaxies as part of the Medium Deep Survey (MDS).
An important scientific aim of the MDS is to identify possible 
AGN candidates from these images in order to measure the faint end of the AGN
luminosity function as well as to study the host galaxies of AGNs and
nuclear starburst systems.  We are able to identify candidate objects
based on morphology.  Candidates are selected by fitting  
bulge+disk models and bulge+disk+point source nuclei models to 
HST imaged galaxies
and determining the best model fit to the galaxy light profile.

We present results from a  
sample of MDS galaxies with I$\lesssim$21.5 mag that have been 
searched for AGN/starburst nuclei in
this manner.  We identify 84 candidates with unresolved nuclei in a 
sample of 825 galaxies.  
For the expected range of galaxy redshifts, all normal bulges are resolved.
Most of the candidates are found in galaxies displaying exponential disks  
with some containing an additional bulge component.  5\% of the hosts
are dominated by an r$^{-1/4}$ bulge.
The V-I color distribution of the nuclei is consistent with a dominant 
population of Seyfert-type nuclei combined with an additional population
of starbursts. Our results suggest that 
$\sim$10\% $\pm$1\% of field galaxies at z$\lesssim$0.6 may 
contain AGN/starburst 
nuclei that are 1 to 5 magnitudes fainter than the host galaxies.  

\end{abstract}

\keywords{galaxies:active-nuclei-starburst}

\section{Introduction}

Accurate knowledge of the luminosity function (LF) of active galactic
nuclei over a wide range of absolute magnitudes is necessary to
understand the nature and evolution of these objects.
The faint end of the AGN LF (M$_B$$\geq$-23) has been 
determined using 
Seyfert galaxy nuclei which are considered to be the
intrinsically fainter counterparts of more distant, brighter AGNs (Cheng
et al. 1985).
The behavior of the LF as a function of redshift can be used to
derive the manner in which AGNs evolve.  
Understanding how the faint end of the AGN LF evolves
is necessary to determine the frequency and total space density of these 
objects at earlier epochs.
One interesting question this information addresses is the 
contribution of Low Luminosity AGNs (LLAGNs)
to the soft X-ray background (2-10 keV) (Elvis et al. 1984)
(Koratkar et al. 1995).

The evolution of the faint end of the LF is difficult to measure. 
Even at modest redshifts, low-luminosity AGNs become virtually
impossible to observe from the ground.  In ground-based images
the unresolved nuclei cannot be distinguished from central regions of
enhanced star formation or finite central density cusps of spheroidal
components.  The Hubble Space Telescope eliminates this problem
with its unique high resolution imaging capabilities. 

The Medium Deep Survey (MDS) (Griffiths et al. 1994) yields 2-4 parallel WFPC2
exposures per week, each containing
$\sim$300 galaxies down to V$\sim$ 23.5 mag.
This survey provides an ideal sample of distant field galaxies for which
morphology and light-profiles can be studied for the first time at sub-kpc
resolution.  Typical galaxy redshifts are within z$\leq$0.6 (Mutz
et al. 1994).  The set of Cycle 4 and
5 images consists of $\sim$100 fields with both V and I
exposures, in which weak stellar nuclei can be distinguished in galaxies
with integrated magnitudes down to
V$\sim$22 mag.  In this analysis, we search all galaxies
in 59 MDS fields to a limiting magnitude 
of I$\lesssim$21.5 for unresolved nuclei
which may indicate a possible AGN or starburst.  

\section{HST Observations}

The HST images used in this study were taken in ``parallel mode'' 
several arcminutes from the primary target in random locations
spanning the entire range of RA and with -16:16$\leq$Dec$\leq$85:20.
Avoiding low galactic latitudes, we chose fields with $\vert$b$\vert$$\geq$20. 
The details of the MDS observing strategy and data reductions can be
found in Griffiths et al. (1994).  On average, there are $\sim$14 galaxies
with I$\leq$21.5 in each WFPC2 field.  We present here 53 MDS fields having 
at least
2 exposures in both V and I and with total exposure times between
2000 and 23,100 seconds.  We also include 6 fields with I images alone having
comparable exposure times. 
Our limiting magnitude of I$\leq$21.5 is at least
3 magnitudes above that for detecting point sources
in any of our selected MDS fields.  This limit allows for uniform galaxy 
detectability across the range of exposure times.

\section{Galaxy Modeling}

We use a modeling approach
based on maximum likelihood estimates to extract
quantitative morphological and structural parameters of the faint
galaxy images in the MDS.  Details of these fitting techniques are
described in Ratnatunga et al. (1994) and Ratnatunga et al. (1996).  We perform
a simultaneous 2-dimensional fit of all major components of 
the light profile: exponential disk, r$^{-1/4}$ bulge, and  
Gaussian point source.  The unconvolved half light radius for a point
source is kept fixed at 0.018$\arcsec$.  
Once this model is convolved with the WFPC2 PSF, the gaussian source 
has a FWHM of $\sim$0.15$\arcsec$ which agrees well with fits to 
stars in the image.
The fitting procedure allows for a choice of models to be fit to each galaxy:
a 7-parameter pure bulge or disk model, an 10-parameter bulge+disk model, 
a 10-parameter disk+point source model,
and a 13-parameter disk+bulge+point source
model.  Chi-square minimization techniques are used
to determine the model parameters of the best fit to the galaxy light
profile.  These parameters include total galaxy magnitude, sky magnitude,
X and Y position,
half-light radius and position angle of the primary component, 
disk and bulge
axis ratios, bulge-to-total galaxy light ratio, 
bulge-to-disk half light radius ratio, point source-to-total 
galaxy light ratio, and position of the point source with respect to 
the disk and/or bulge center.   The point source position was constrained to be
$\lesssim$0.2$\arcsec$ from the disk or bulge central position to avoid 
fitting non-nuclear knots and irregularities
seen in many galaxies with active star formation.
The best model fit is determined by comparing
the computed likelihood estimation values.
In addition each galaxy was individually examined to verify the goodness
of fit for each model and note highly irregular galaxies which were not
well fit by any model.

It is important to find not only the best fit, but a unique fit to each
galaxy.  In some cases a fit with the point source included was not
significantly better than the simple bulge+disk model alone.  To determine 
the significance of the difference in computed likelihood estimates of
each fit, we performed the same fitting procedures on simulated galaxies
with a range in magnitude, bulge-to-total light ratio, and half-light radius.
These galaxies did not contain point source nuclei and consisted of pure
bulge+disk components.  We also conducted fitting tests using real galaxy 
images from our sample with simulated point sources superimposed 
in the nucleus.  Based on the results of the fits to these simulated
galaxies, we were able to determine a significance level for the detection of a
point source component in our sample galaxies.  This significance level was 
used to filter spurious detections
of point source components which were insignificant as compared to the
simple bulge+disk model fits for our galaxies. 

An unresolved source at z=0.1, assuming WFPC2 resolution
of 0.15$\arcsec$, corresponds to an actual size of $<$370 pc, using H$_o$=75
km/s/Mpc and q$_o$=0.5.  At z=0.6, this corresponds to a size of $<$1140 pc.
The emitting regions of Seyfert 1 or Seyfert 2 nuclei as well as those of 
smaller starburst regions, would lie well within these boundaries.
These sizes are smaller than the typical bulge sizes of late type spirals
making it unlikely that the 
unresolved regions in the majority of our galaxies are due 
to a simple bulge component.  Our
most distant late-type spirals, however, could contain bulge components 
that are near the unresolved region size and may therefore be
included in our sample.  The contribution of unresolved bulges should
be small since the current galaxy redshift distribution for MDS galaxies 
peaks near z$\sim$0.3 and steadily decreases to z$\sim$0.6.

\section{Number Counts}

We find 84 galaxies requiring a nuclear, unresolved point source component 
in the galaxy model.  This sample consists of
21 sources detected in I frames only, 8 in V frames only,
and 55 in both I and
V representing 10.2$\pm$1.1\% of the total 825 galaxies in our
magnitude-limited survey.  Almost all of the host galaxies for these sources
contained significant exponential disks,
consistent with the morphology of Seyfert galaxies or nuclear starburst 
galaxies.
57\% of the host galaxies were adequately modeled with an exponential disk 
alone
while 38\% required an additional bulge component in the fit.  5\% required
a dominant bulge component where the bulge contributes $\ge$80\% of the
total galaxy light. 
We note that our search is not sensitive to pure stellar-like
quasars which would probably be identified as stars in our initial galaxy
search.   Apparent magnitudes for the nuclei range from 20$\leq$I$\leq$26
and 21$\leq$V$\leq$27.
If we assume our galaxies to be at
0.1$\le$z$\le$0.6, a probable absolute magnitude 
range for our nuclei is
-21$\lesssim$M$_{V}$$\lesssim$-14.  The nuclei make up between 1\%
and 50\% of the total galaxy light with all but 8 having nuclei
comprising less than 20\%.  Detections to the 1\% level 
were obtained in all fields across the range of exposure times such
that there is no apparent incompleteness due to short exposure time fields. 
Figure 1 shows examples of
galaxies in our sample requiring a nuclear point source component in
the galaxy fit. 

Of the remaining 741 galaxies in our magnitude-limited sample not
requiring an additional point source component,
463 required a bulge and disk fit in both V and I to accurately model
the galaxy light.  160 required a pure disk model and 100 required a pure
or dominating ($\geq$80\%) bulge model. 18 were not well fit by any model 
due to irregularity.

We estimate our incompleteness by comparing the measured 
nuclear magnitude to that of the total integrated galaxy magnitude.  
We find that the faintest point source component which can be detected
comprises $\sim$1\% of the total galaxy light.  
At a galaxy limiting magnitude of I$\leq$21.5, we are above the
photon noise limited regime in detecting point source components.
Our detection limit is imposed by noise elements in the images such
as flat fielding effects, cosmic ray removal, etc.
Some galaxies contained additional structural
elements in the nucleus such as rings and bars,
but they were found to be relatively insignificant noise sources. 
Based on our detection limits and the distribution of galaxy magnitudes
in our sample,
we are $\sim$80\% complete in searching for nuclei
down to 5 mags fainter than the total galaxy apparent magnitude
for all galaxies in both V and I.  For galaxies at the bright end
of our distribution (19.5$\lesssim$I$\lesssim$18.0) the detection limit
for a point source component increases from 1\% to $\sim$3\% of the 
total galaxy light.  Small number statistics at the bright end of our
distribution make it difficult to determine the completeness in searching
for faint nuclei in these galaxies.  We estimate that we are $\sim$93\%
complete in searching for nuclei down to 3.8 mags fainter than the galaxy
apparent magnitude.  We have not incorporated this incompleteness into
our number counts of unresolved nuclei.

\section{Nuclei and Host Galaxy Colors}

Figure 2 shows the normalized distribution of V-I colors
for the 55 nuclei in our
sample where accurate V {\it and} I magnitudes were measurable. 
The typical error in V-I color is $\pm$$\sim$0.2 mag.   
The mean V-I color of known Seyfert 1s/QSOs
from Elvis et al. (1994) at z$\leq$0.5,
typical of the galaxy redshifts in MDS fields, is indicated at 0.85.
These QSOs/Seyfert 1s are  
dominated in ground-based images by their active nuclei and 
should be valid for comparison with our nuclear colors.  
We also indicate the mean V-I color of Seyfert 2 nuclei based on
data from Kalinkov et al. (1993).  They measure a mean B-V color
for Seyfert 2 nuclei of 0.71.  Assuming a power-law spectral energy 
distribution, we transform to V-I and estimate a color of 1.35.   

Our sample reveals a large range in colors with a dominant population
at V-I$\simeq$1.25.  This distribution might be explained by the
presence of more than one object type in our sample.
A broad, shallow distribution may result from starbursts 
seen at various stages of evolution
and/or broadened by reddening due to dust.  
Colors for starburst nuclei based on spectra from Bica et al (1990) 
range from V-I = 0.2 during the young HII region phase to V-I = 1.3 or more 
for intermediate age clusters at $\ge$ 7 Gyr.
The blue tail of our distribution may consist of starburst nuclei
since the colors are bluer than those expected
for QSOs/Sy1s and are probably due to the presence of young stars.
Our peak falls close to the predicted Sy2 nuclear color
with a blue side asymmetry biased toward Sy1
colors suggesting that the dominant population may be AGN-related.

Figure 3 shows the integrated host galaxy color versus the nuclear
color in V-I for the 55 sources detected in both the V and I images.
The host galaxies have colors in the range 
0.5$\lesssim$V-I$\lesssim$2 
with roughly half having bluer nuclei and half containing redder nuclei.
There is not a 1-to-1 relation between the galaxy color and nuclear color,
indicating that the point source is not simply a fitting residual in the
galaxy.  We did not require that the nuclei necessarily differ in color from
the underlying host galaxy for sample selection.  
However, an elliptical or bulge-dominated
galaxy could be sufficiently cuspy in the nucleus such that an additional 
point source component improves the model fit.  To avoid cuspy bulges in our 
selected sample, we removed 8 galaxies
which were well fit with an elliptical-type galaxy plus a
point source of the same color as the galaxy.

\section{Summary and Discussion}

In summary, we find that in a magnitude-limited
survey (I$\leq$21.5) of 825 galaxies estimated to be at moderate redshifts,
84 contain unresolved nuclear point sources.
This represents 10.2$\pm$1.1\% of all galaxies
in our sample.  Almost all of the host galaxies of these nuclei are spirals
with some displaying significant bulge components and a small fraction
residing in elliptical hosts. 

V-I colors for the nuclei cover a large range and might be
attributed to the presence of different object types in our sample. The
distribution is consistent with a 
population of Seyfert-like galaxies combined with a population of 
starburst nuclei
at various stages in their evolution or which have been reddened to various
degrees by the presence of dust.

Since the V-I colors of starburst nuclei and AGN nuclei are likely to
overlap, we cannot explicitly separate the two populations in our sample.
Due to the peaked nature of our color distribution, we might assume that
one or the other dominates.  For the purposes of comparison, 
let us assume that the dominant population represented in our nuclei sample
is AGN-like.  If so, how do our counts
compare with previous studies of the 
incidence of these objects?  Recent results from the Canada-France Redshift
Survey (Tresse et al. 1996) find that between 8\% and
17\% of the galaxies at z$\leq$0.3 are narrow-line
AGNs based on line ratio diagnostics. 
Locally, Huchra and Burg (1992) find the fraction of AGNs in
the CfA redshift survey of 2400 galaxies
to be $\sim$2\% based on spectroscopic identification.
Differences in identification of AGN based on spectroscopy and
morphology, however, allow
intrinsically fainter AGNs to be detected in our survey.
Although our galaxy absolute magnitudes may cover the same range,
our nuclear absolute magnitudes are certainly fainter than those observed
in the CfA or CFRS studies.
Granato et al. (1993) show that many Seyfert 1s in the CfA
survey have nuclei contributing between 30\% and 90\% of the total galaxy light
whereas our nuclei typically contribute less than 20\%.
A fainter absolute magnitude limit combined
with the apparent steepness of the AGN LF at low redshifts (Cheng et al. 1985) 
could account for an apparent increase in number density from 2\% locally
to $\sim$10\% at moderate redshifts.  

We are currently obtaining ground-based spectra of galaxies containing 
AGN/starburst nuclear candidates for redshift information 
as well as possible spectral
identifications.  Also, recent MDS fields include B-band exposures in
addition to V and I which will provide increased color information
for the identified nuclei.  We will be expanding this study to include
more MDS fields to increase our sample of unresolved nuclei.    
With the additional spectroscopic and color
information, the question of LF evolution of fainter AGNs can be
addressed directly. 

\acknowledgments
We would like to thank the referee, Jonathan McDowell, for his useful
suggestions for improving this paper.
This work is based on observations taken with the NASA/ESA Hubble
Space Telescope, obtained at the Space Telescope Science Institute,
operated by the Associations of Universities for Research in Astronomy,
Inc.  This work was supported in part by STScI Grant GO-02684.06-87A
to RFG for the Medium Deep Survey project.

\clearpage

\centerline{FIGURE CAPTIONS}

Figure 1 - V filter image of MDS galaxies containing unresolved nuclear 
components. 

Figure 2 - (V-I) color histogram for unresolved nuclear sources. 
The mean (V-I) color of 49 Seyfert 1s/QSOs from Elvis et al. (1994)
is indicated by an arrow as well as the mean color for Seyfert 2s
based on data from Kalinkov et al. (1993).

Figure 3 - (V-I) color of the integrated galaxy versus the (V-I)
color of the unresolved nucleus for the 55 sources detected in
both the V and I images.
  
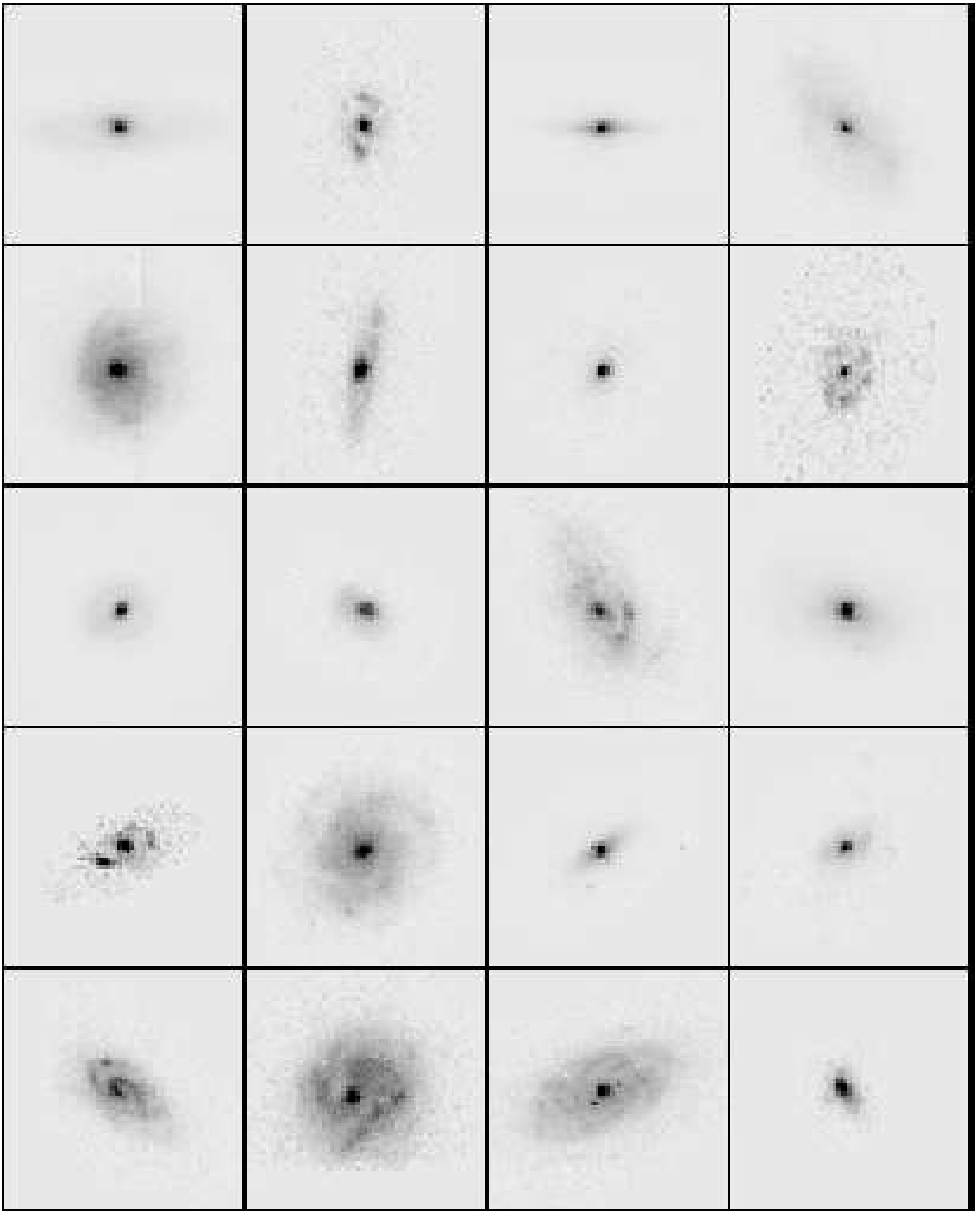
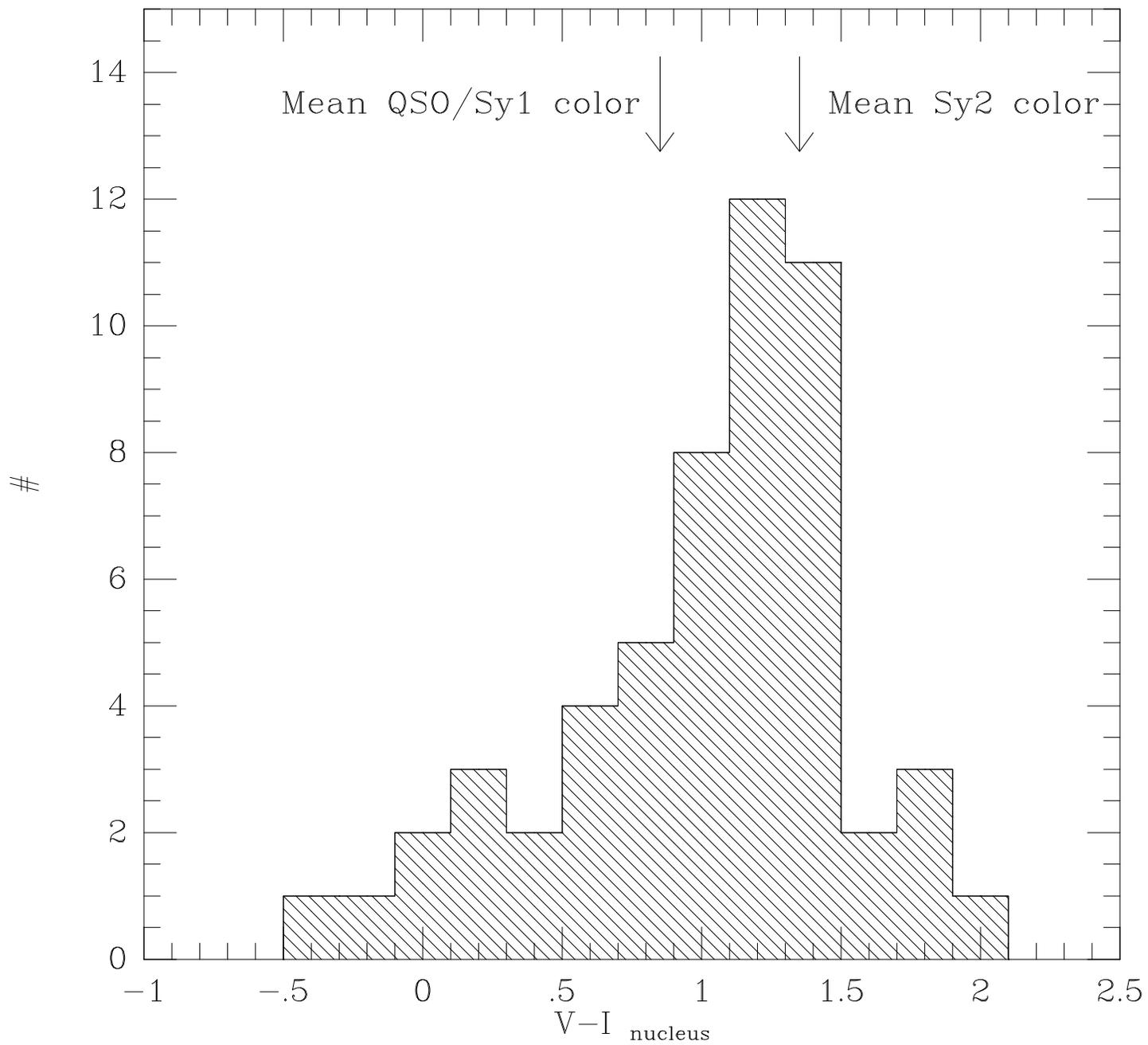
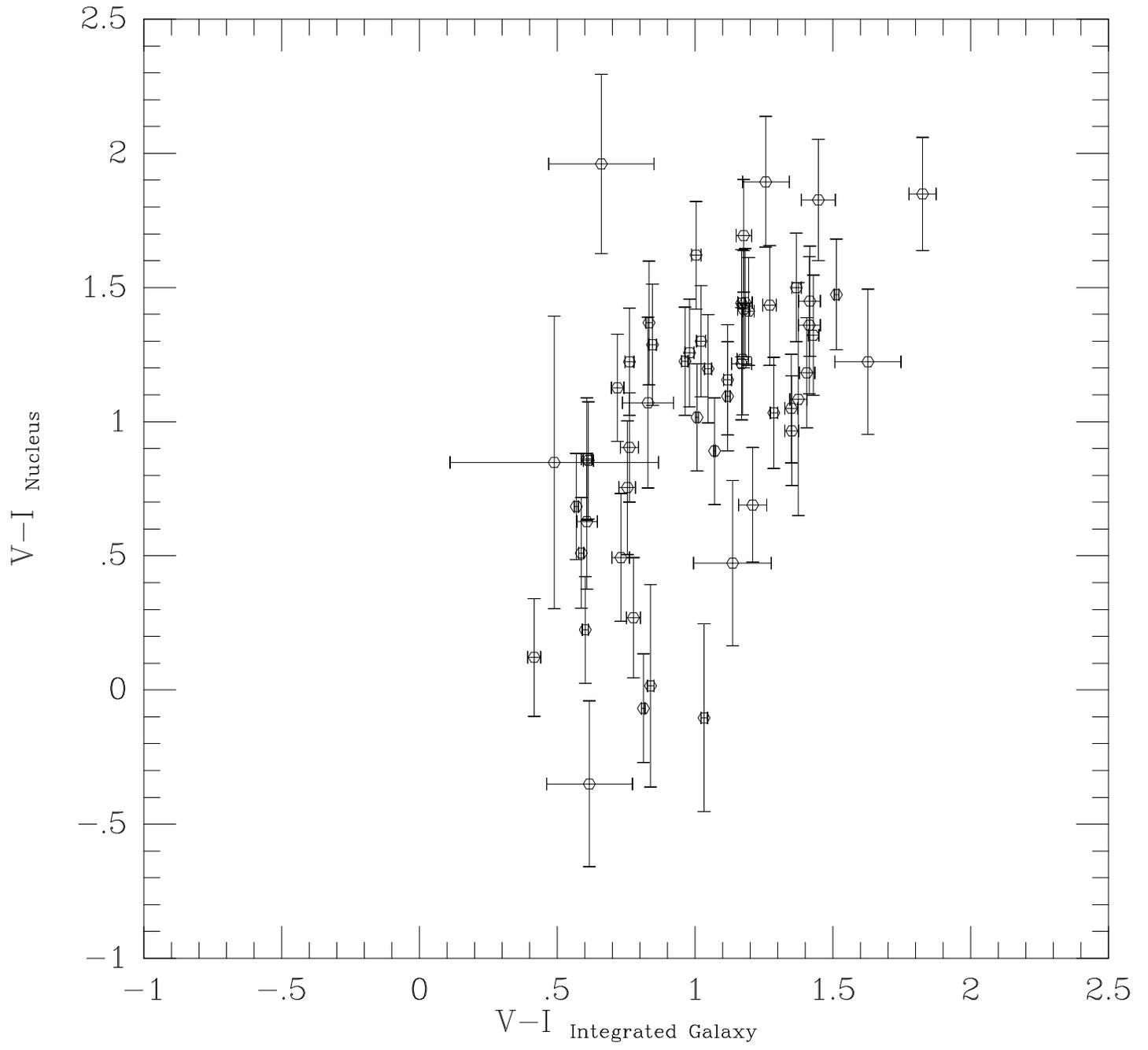
\end{document}